\documentclass[12pt,onecolumn]{revtex4}
\usepackage{amssymb}

\begin{document}

\title{Magnetic Branes in Gauss-Bonnet Gravity}
\author{M. H. Dehghani}\email{dehghani@physics.susc.ac.ir}
\address{Physics Department and Biruni  Observatory,
         Shiraz University, Shiraz 71454, Iran\\ and\\
         Institute for Studies in Theoretical Physics and Mathematics (IPM)\\
         P.O. Box 19395-5531, Tehran, Iran}

\begin{abstract}
We present two new classes of magnetic brane solutions in
Einstein-Maxwell-Gauss-Bonnet gravity with a negative cosmological
constant. The first class of solutions yields an
$(n+1)$-dimensional spacetime with a longitudinal magnetic field
generated by a static magnetic brane. We also generalize this
solution to the case of spinning magnetic branes with one or more
rotation parameters. We find that these solutions have no
curvature singularity and no horizons, but have a conic geometry.
In these spacetimes, when all the rotation parameters are zero,
the electric field vanishes, and therefore the brane has no net
electric charge. For the spinning brane, when one or more rotation
parameters are non zero, the brane has a net electric charge which
is proportional to the magnitude of the rotation parameter. The
second class of solutions yields a spacetime with an angular
magnetic field. These solutions have no curvature singularity, no
horizon, and no conical singularity. Again we find that the net
electric charge of the branes in these spacetimes is proportional
to the magnitude of the velocity of the brane. Finally, we use the
counterterm method in the Gauss-Bonnet gravity and compute the
conserved quantities of these spacetimes.
\end{abstract}

\maketitle


\section{Introduction}

The possibility that spacetime may have more than four dimensions
is now a standard assumption in high energy physics. From a
cosmological point of view, our observable Universe may be viewed
as a brane embedded into a higher dimensional spacetime. In the
context of string theory, extra dimensions were promoted from an
interesting curiosity to a theoretical necessity since superstring
theory requires a ten-dimensional spacetime to be consistent from
the quantum point of view. The idea of brane cosmology is also
consistent with string theory, which suggests that matter and
gauge interaction (described by an open string) may be localized
on a brane, embedded into a higher dimensional spacetime. The
field represented by closed strings, in particular, gravity,
propagate in the whole of spacetime.

This underscores the need to consider gravity in higher
dimensions. In this context one may use another consistent theory
of gravity in any dimension with a more general action. This
action may be written, for example, through the use of string
theory. The effect of string theory on classical gravitational
physics is usually investigated by means of a low energy effective
action which describes gravity at the classical level \cite{Wit}.
This effective action consists of the Einstein-Hilbert action plus
curvature-squared terms and higher powers as well, and in general
give rise to fourth order field equations and bring in ghosts.
However, if the effective action contains the higher powers of
curvature in particular combinations, then only second order field
equations are produced and consequently no ghosts arise \cite
{Zum}. The effective action obtained by this argument is precisely
of the form proposed by Lovelock \cite{Lov}. The appearance of
higher derivative gravitational terms can be seen also in the
renormalization of quantum field theory in curved spacetime
\cite{BDav}.

In this paper we want to restrict ourself to the first three terms
of Lovelock gravity. The first two terms are the Einstein-Hilbert
term with cosmological constant, while the third term is known as
the Gauss-Bonnet term. This term appears naturally in the
next-to-leading order term of the heterotic string effective
action and plays a fundamental role in Chern-Simons gravitational
theories \cite{Cham}. From a geometric point of view, the
combination of the Einstein-Gauss-Bonnet terms constitutes, for
five-dimensional spacetimes, the most general Lagrangian producing
second order field equations, as in the four-dimensional gravity
where the Einstein-Hilbert action is the most general Lagrangian
producing second order field equations \cite{Lan}.

These facts provide a strong motivation for considering new exact
solutions of the Einstein-Gauss-Bonnet gravity. Because of the
nonlinearity of the field equations, it is very difficult to find
out nontrivial exact analytical solutions of Einstein's equation
with higher curvature terms. In most cases, one has to adopt some
approximation methods or find solutions numerically. However,
static spherically symmetric black hole solutions of the
Gauss-Bonnet gravity were found in Ref. \cite{Des}. Black hole
solutions with nontrivial topology in this theory were also
studied in Refs. \cite{Cai,Aros,Ish}. The thermodynamics of
charged static spherically symmetric black hole solutions was
considered in \cite{Odin}. All of these known solutions are
static. Recently I introduced a new class of asymptotically
anti-de Sitter rotating black brane solutions in the
Einstein-Gauss-Bonnet gravity and considered its thermodynamics
\cite{Deh1}.

In this paper we are dealing with the issue of the spacetimes
generated by static, spinning, and traveling brane sources in
$(n+1)$-dimensional Einstein-Maxwell-Gauss-Bonnet gravity that are
horizonless and have nontrivial external solutions. These kinds of
solutions have been investigated by many authors in the context of
Einstein gravity. Static uncharged cylindrically symmetric
solutions of Einstein gravity in four dimensions were considered
in \cite{Levi}. Similar static solutions in the context of cosmic
string theory were found in \cite{Vil}. All of these solutions
\cite{Levi,Vil} are horizonless and have a conical geometry, which
are everywhere flat except at the location of the line source. An
extension to include the electromagnetic field has also been done \cite{Muk,Lem1}%
. The generalization of the four-dimensional solution found in
\cite{Lem1} to the case of $(n+1)$-dimensional solution with all
rotation and boost parameters has been done in \cite{Deh2}. Some
solutions of type IIB supergravity compactified on a
four-dimensional torus have been considered in \cite{Lun}, which
have no curvature singularity and no conic singularity. Here we
will find these kinds of solutions in the Gauss-Bonnet gravity,
and use the counterterm method to compute the conserved quantities
of the system.

The outline of our paper is as follows. We give a brief review of
the field equations in Sec. \ref{Fiel}. In Sec. \ref{Lon} we first
present a new class of static horizonless solutions which produce
longitudinal magnetic field, and then endow these spacetime
solutions with a global rotation. We also generalize these
rotating solutions to the case of spacetimes with more rotation
parameters. In Sec. \ref{AngMag} we introduce those horizonless
solutions that produce an angular magnetic field. Section
\ref{Conserv} will be devoted to the use of the counterterm method
to compute the conserved quantities of these spacetimes. We also
show that the electric charge densities of the branes with
rotation or a boost parameter are proportional to the magnitude of
the boost parameters. We finish our paper with some concluding
remarks.

\section{Field Equations in Einstein-Maxwell-Gauss-Bonnet Gravity}

\label{Fiel}

The most fundamental assumption in standard general relativity are
the requirement of general covariance and that the field equations
be second order. Based on these principles, the most general
Lagrangian in arbitrary dimensions is the Lovelock Lagrangian. The
Lagrangian of the Lovelock theory, which is the sum of
dimensionally extended Euler densities, may be written as
\begin{equation}
\mathcal{L}_{G}=\frac{1}{2} \sum_{i}^{n}c_{i}\mathcal{L}_{i},  \label{Lov1}
\end{equation}
where $c_{i}$ is an arbitrary constant and $\mathcal{L}_{i}$ is the Euler
density of a $2i$-dimensional manifold,
\begin{equation}
\mathcal{L}_{i}=(-2)^{-i}\delta
_{c_{1}d_{1}...c_{i}d_{i}}^{a_{1}b_{1}...a_{i}b_{i}}R_{a_{1}b_{1}}^{c_{1}d_{1}}....R_{a_{i}b_{i}}^{c_{i}d_{i}}.
\label{Lov2}
\end{equation}
Here the first two terms $c_{0}\mathcal{L}_{0}=2\Lambda $, where $\Lambda $
is the cosmological constant, and $c_{1}\mathcal{L}_{1}=R$ give us the
Einstein-Hilbert term and $c_{2}\mathcal{L}_{2}=\alpha (R_{\mu \nu \gamma
\delta }R^{\mu \nu \gamma \delta }-4R_{\mu \nu }R^{\mu \nu }+R^{2})$ is the
Gauss-Bonnet term. Thus, the gravitational action of an $(n+1)$-dimensional
asymptotically anti-de Sitter spacetimes with the Gauss-Bonnet term in the
presence of an electromagnetic field, in a unit system in which $8\pi G=1$,
can be written as
\begin{equation}
I_{g}=\frac{1}{2}\int d^{n+1}x\sqrt{-g}\{R+\frac{n(n-1)}{l^{2}}+\alpha
(R_{\mu \nu \gamma \delta }R^{\mu \nu \gamma \delta }-4R_{\mu \nu }R^{\mu
\nu }+R^{2})+F_{\mu \nu }F^{\mu \nu }\},  \label{Actg}
\end{equation}
where $R$, $R_{\mu\nu\rho\sigma}$, and $R_{\mu\nu}$ are the Ricci
scalar and Riemann and Ricci tensors of the spacetime, $F_{\mu \nu
}=\partial _{\mu }A_{\nu }-\partial _{\nu }A_{\mu }$ is the
electromagnetic tensor field, and $A_{\mu }$ is the vector
potential. $\alpha $ is the Gauss-Bonnet coefficient with
dimension $(\emph{length})^{2}$ and is positive in the heterotic
string theory \cite{Des}. So we restrict ourselves to the case
$\alpha \geq 0 $. Varying the action over the metric tensor
$g_{\mu \nu }$ and electromagnetic field $F_{\mu \nu }$, the
equations of gravitational and electromagnetic fields are obtained
as
\begin{eqnarray}
&&\ R_{\mu \nu }-\frac{1}{2}g_{\mu \nu }R+\frac{n(n-1)}{2l^{2}}g_{\mu \nu
}-\alpha \{\frac{1}{2}g_{\mu \nu }(R_{\gamma \delta \lambda \sigma
}R^{\gamma \delta \lambda \sigma }-4R_{\gamma \delta }R^{\gamma \delta
}+R^{2})  \nonumber \\
&&-2RR_{\mu \nu }+4R_{\mu \gamma }R_{\ \nu }^{\gamma }+4R_{\gamma \delta
}R_{\mu \nu }^{\gamma \ \delta }-2R_{\mu \gamma \delta \lambda }R_{\nu }^{\
\gamma \delta \lambda }\}=T_{\mu \nu },  \label{Geq}
\end{eqnarray}
\begin{equation}
\nabla _{\mu }F_{\mu \nu }=0,  \label{EMeq}
\end{equation}
where $T_{\mu \nu }$ is the electromagnetic stress tensor
\begin{equation}
T_{\mu \nu }=2{F^{\lambda }\ _{\mu }}F_{\lambda \nu }-\frac{1}{2}F_{\lambda
\sigma }F^{\lambda \sigma }g_{\mu \nu }.  \label{Str}
\end{equation}
Equation (\ref{Geq}) does not contain the derivative of the curvatures, and
therefore the derivatives of the metric higher than two do not appear. Thus,
the Gauss-Bonnet gravity is a special case of higher derivative gravity.

\section{The Longitudinal Magnetic Field Solutions}

\label{Lon}

Here we want to obtain the $(n+1)$-dimensional solutions of Eqs. (\ref{Geq}%
)-(\ref{Str}) which produce longitudinal magnetic fields in the
Euclidean submanifold spans by the $x^{i}$ coordinates
($i=1,...,n-2$). We assume that the metric has the following form:
\begin{equation}
ds^{2}=-\frac{\rho ^{2}}{l^{2}}dt^{2}+\frac{d\rho ^{2}}{f(\rho )}%
+l^{2}f(\rho )d\phi ^{2}+\frac{\rho ^{2}}{l^{2}}{{\sum_{i=1}^{n-2}}}%
(dx^{i})^{2}.  \label{Met1a}
\end{equation}
Note that the coordinates $x^{i}$ have the dimension of length, while the
angular coordinate $\phi $ is dimensionless as usual and ranges in $%
0\leq \phi <2 \pi$. The motivation for this metric gauge $[g_{tt}\varpropto
-\rho ^{2}$ and $(g_{\rho \rho })^{-1}\varpropto g_{\phi \phi }]$ instead of
the usual Schwarzschild gauge $[(g_{\rho \rho })^{-1}\varpropto g_{tt}$ and $%
g_{\phi \phi }\varpropto \rho ^{2}]$ comes from the fact that we
are looking for a magnetic solution instead of an electric one. In
this section we want to consider only the magnetically charged
case which produces a longitudinal magnetic field in the Euclidean
submanifold spanned by the $x^{i}$ coordinates. In the next
section we consider the angular magnetic solutions. Thus, one may
assume that
\begin{equation}
A_{\mu }=h(\rho )\delta _{\mu }^{\phi }.  \label{Pot1a}
\end{equation}
For purely electrically charged solutions in Gauss-Bonnet gravity,
see \cite {Deh1}. The functions $f(\rho )$ and $h(\rho )$ may be
obtained by solving the field equations (\ref{Geq}) and
(\ref{EMeq}). Using Eq. (\ref{EMeq}) one obtains
\begin{equation}
\rho \frac{\partial ^{2}h}{\partial \rho ^{2}}+(n-1)\frac{\partial h}{%
\partial \rho }=0.  \label{Em2}
\end{equation}
Thus, $h(\rho )=C_{1}/\rho ^{(n-2)}$, where $C_{1}$ is an arbitrary real
constant. To get the solution of the Einstein-Maxwell equation in the case of $%
\alpha =0$, which I introduced in \cite{Deh2}, one should choose
the
arbitrary constant $C_{1}=-2ql^{n-1}/(n-2)$. To find the function $f(\rho )$%
, one may use any components of Eq. (\ref{Geq}). The simplest
equation is the $\rho \rho$ component of these equations which can
be written as
\begin{eqnarray}
&&(n-1)\{l^{2}\rho^{2n-5}[2(n-2)(n-3)\alpha f-\rho ^{2}]f^{^{\prime }}+n\rho
^{2n-2}  \nonumber \\
&&+(n-2)l^{2}\rho ^{2n-6}[(n-3)(n-4)\alpha f-\rho ^{2}]f\} +8q^{2}l^{2n-2}
=0,  \label{rrcom}
\end{eqnarray}
where the prime denotes a derivative with respect to the $\rho $ coordinate.
The solutions of Eq. (\ref{rrcom}) can be written as
\begin{equation}
f(\rho )=\frac{\rho ^{2}}{2(n-2)(n-3)\alpha }\left\{ 1\pm \sqrt{1-\frac{%
4(n-2)(n-3)\alpha}{l^2} \left( 1+\frac{C_{2}}{\rho ^{n}}-\frac{8q^{2}l^{2n-2}%
}{(n-1)(n-2)\rho ^{2(n-1)}}\right) }\right\},  \label{Fg1}
\end{equation}
where $C_{2}$ is an arbitrary constant. As one can see from Eq.
(\ref{Fg1}), the solution has two branches with ``$-$'' and
``$+$'' signs. The arbitrary constant $C_{2}$ should be chosen
such that the solution obtained reduces to that of the
Einstein-Maxwell equation introduced in \cite{Deh2} as $\alpha $
goes to zero. In order to have the desired function, we should
choose the branch with the ``$-$'' sign and fix $C_{2}=-8ml^n$.
The parameters $m$ and $q$ are the mass and charge parameters and
$r_{+}$ is the largest positive real solution of $f(\rho)=0$.

In order to study the general structure of this solution, we first look for
curvature singularities. It is easy to show that the Kretschmann scalar $%
R_{\mu \nu \lambda \kappa }R^{\mu \nu \lambda \kappa }$ diverges
at $\rho =0$ and therefore one might think that there is a
curvature singularity located at $\rho =0$. However, as will be
seen below, the spacetime will never achieve $\rho =0$. Now we
look for the existence of horizons, and therefore we look for
possible black brane solutions. One should conclude that there are
no horizons and therefore no black brane solutions. The horizons,
if any exist, are given by the zeros of the function $f(\rho
)=g_{\rho \rho }^{-1}$. Let us denote the zeros of $f(\rho )$ by
$r_{+}$. The function $f(\rho )$ is negative for $\rho <r_{+}$ and
positive for $\rho >r_{+}$, and therefore one may think that the
hypersurface of constant time and $\rho =r_{+}$ is the horizon.
However, this analysis is not correct. Indeed, one may note that
$g_{\rho \rho }$ and $g_{\phi \phi }$ are related by $f(\rho
)=g_{\rho \rho }^{-1}=l^{-2}g_{\phi \phi }$, and therefore when
$g_{\rho \rho }$ becomes negative (which occurs for $\rho <r_{+}$)
so does $g_{\phi
\phi }$. This leads to an apparent change of signature of the metric from $%
(n-1)+$ to $(n-2)+$, and therefore indicates that we are using an incorrect
extension. To get rid of this incorrect extension, we introduce the new
radial coordinate $r$ as
\begin{equation}
r^{2}=\rho ^{2}-r_{+}^{2}\Rightarrow d\rho ^{2}=\frac{r^{2}}{r^{2}+r_{+}^{2}}%
dr^{2}.
\end{equation}
With this new coordinate, the metric (\ref{Met1a}) is
\begin{eqnarray}
ds^{2} &=&-\frac{r^{2}+r_{+}^{2}}{l^{2}}dt^{2}+l^{2}f(r)d\phi ^{2}  \nonumber
\\
&&+\frac{r^{2}}{(r^{2}+r_{+}^{2})f(r)}dr^{2}+\frac{r^{2}+r_{+}^{2}}{l^{2}}%
dX^{2},  \label{Met1b}
\end{eqnarray}
where $dX^{2}$ is the Euclidean metric on the $(n-2)$-dimensional
submanifold,
the coordinates $r$ and $\phi $ assume the values $0\leq r<\infty $ and $%
0\leq \phi <2\pi $, and $f(r)$ is now given as
\begin{eqnarray}
&&f(r)=\frac{r^{2}+r_{+}^{2}}{2(n-2)(n-3)\alpha }\{1-  \nonumber \\
&&\hspace{0.5cm}\sqrt{1-\frac{4(n-2)(n-3)\alpha }{l^{2}}\left( 1-\frac{%
8ml^{n}}{(r^{2}+r_{+}^{2})^{n/2}}-\frac{8q^{2}l^{2n-2}}{%
(n-1)(n-2)(r^{2}+r_{+}^{2})^{n-1}}\right) }\}.  \label{Fg2}
\end{eqnarray}
The gauge potential in the new coordinate is
\begin{equation}
A_{\mu }=-\frac{2}{(n-2)}\frac{ql^{n-1}}{(r^{2}+r_{+}^{2})^{(n-2)/2}}\delta
_{\mu }^{\phi }.  \label{Pot1b}
\end{equation}
The function $f(r)$ given in Eq. (\ref{Fg2}) is positive in the whole
spacetime and is zero at $r=0$. Also note that the Kretschmann scalar does
not diverge in the range $0\leq r<\infty $. Therefore this spacetime has no
curvature singularities and no horizons. However, it has a conic geometry
and has a conical singularity at $r=0$. In fact, using a Taylor expansion,
in the vicinity of $r=0$ the metric (\ref{Met1b}) is
\begin{eqnarray}
ds^{2} &=&-\frac{r_{+}^{2}}{l^{2}}dt^{2}+2\frac{l^{2}}{r_{+}^{2}}\left( n+%
\frac{4q^{2}l^{2n-2}}{(n-1)r_{+}^{2n-2}}\right) ^{-1}dr^{2}  \nonumber \\
&&+\frac{1}{2l^{2}}\left( n+\frac{4q^{2}l^{2n-4}}{(n-1)r_{+}^{2n-2}}\right)
r^{2}d\phi^{2}+\frac{r_{+}^{2}}{l^2}dX^{2},
\end{eqnarray}
which clearly shows that the spacetime has a conical singularity at $r=0$.
It is worthwhile to mention that the magnetic solutions obtained here have
distinct properties relative to the electric solutions obtained in \cite
{Deh1}. Indeed, the electric solutions have black holes, while the magnetic
do not.

Of course, one may ask for the completeness of the spacetime with
$r\geq 0$ \cite{Lem1,Hor}. It is easy to see that the spacetime
described by Eq. (\ref {Met1b}) is both null and timelike
geodesically complete for $r\geq 0$. To do this, one may show that
every null or timelike geodesic starting from an arbitrary point
either can be extended to infinite values of the affine parameter
along the geodesic or will end on a singularity at $r=0$. Using
the geodesic equation, one obtains
\begin{eqnarray}
&&\dot{t}=\frac{l^{2}}{r^{2}+r_{+}^{2}}E,\hspace{0.5cm}\dot{x^{i}}=\frac{%
l^{2}}{r^{2}+r_{+}^{2}}P^{i},\hspace{0.5cm}\dot{\phi}=\frac{1}{l^{2}f(r)}L,
\label{Geo1} \\
&&r^{2}\dot{r}^{2}=(r^{2}+r_{+}^{2})f(r)\left[ \frac{l^{2}(E^{2}-\mathbf{P}%
^{2})}{r^{2}+r_{+}^{2}}-\alpha \right] -\frac{r^{2}+r_{+}^{2}}{l^{2}}L^{2},
\label{Geo2}
\end{eqnarray}
where the overdot denotes the derivative with respect to an affine
parameter, and
$\alpha $ is zero for null geodesics and $+1$ for timelike geodesics. $E$, $%
L$, and $P^{i}$'s are the conserved quantities associated with the
coordinates $t$, $\phi$, and $x^{i}$'s, respectively, and $\mathbf{P}%
^{2}=\sum_{i=1}^{n-2}(P^{i})^{2}$. Notice that $f(r)$ is always positive for
$r>0$ and zero for $r=0$.

First we consider the null geodesics ($\alpha=0$). (i) If $E^{2}>\mathbf{P}%
^{2}$ the spiraling particles ($L>0$) coming from infinity have a
turning point at $r_{tp}>0$, while the nonspiraling particles
($L=0$) have a turning point at $r_{tp}=0$. (ii) If $E=\mathbf{P}$
and $L=0$, whatever the value of $r$, $\dot{r}$ and $\dot{\phi}$
vanish and therefore the null particles moves in a straight line
in the $(n-2)$-dimensional submanifold spanned by $x^{1}$ to
$x^{n-2}$. (iii) For $E=\mathbf{P}$ and $L\neq0$, and also for
$E^{2}<\mathbf{P}^{2}$ and any values of $L$, there is no possible
null geodesic.

Now, we analyze the timelike geodesics ($\alpha =+1$). Timelike geodesics is
possible only if $l^{2}(E^{2}-\mathbf{P}^{2})>r_{+}^{2}$. In this case
spiraling ($L\neq 0$) timelike particles are bound between $r_{tp}^{a}$ and $%
r_{tp}^{b}$ given by
\begin{equation}
0<r_{tp}^{a}\leq r_{tp}^{b}<\sqrt{l^{2}(E^{2}-\mathbf{P}^{2})-r_{+}^{2}},
\end{equation}
while the turning points for the nonspiraling particles ($L=0$) are $%
r_{tp}^{1}=0$ and $r_{tp}^{2}=\sqrt{l^{2}(E^{2}-\mathbf{P}^{2})-r_{+}^{2}}$.
Thus, we confirmed that the spacetime described by Eq. (\ref{Met1b}) is both
null and timelike geodesically complete.

When $m$ and $q$ are zero, the vacuum solution is
\begin{equation}
f(\rho )=\frac{\rho ^{2}}{2(n-2)(n-3)\alpha }\left( 1-\sqrt{1-\alpha \frac{%
4(n-2)(n-3)}{l^{2}}}\right) .  \label{Fg0}
\end{equation}
Equation (\ref{Fg0}) shows that for a positive value of $\alpha $, this
parameter should be less than $\alpha \leq l^{2}/4(n-2)(n-3)$. Also note
that the AdS solution of the theory has the effective cosmological constant
\begin{equation}
l_{\mathrm{eff}}^{2}=2(n-2)(n-3)\alpha \left( 1-\sqrt{1-\frac{%
4(n-2)(n-3)\alpha }{l^{2}}}\right) ^{-1}.  \label{leff}
\end{equation}
We use the effective cosmological constant (\ref{leff}) later in order to
introduce the counterterm for the action.

\subsection{The rotating longitudinal solutions}

\label{Angul} Now, we want to endow our spacetime solution (\ref{Met1b})
with a global rotation. In order to add angular momentum to the spacetime,
we perform the following rotation boost in the $t-\phi $ plane
\begin{equation}
t\mapsto \Xi t-a\phi ,\hspace{0.5cm}\phi \mapsto \Xi \phi -\frac{a}{l^{2}}t,
\label{Tr}
\end{equation}
where $a$ is a rotation parameter and $\Xi =\sqrt{1+a^{2}/l^{2}}$.
Substituting Eq. (\ref{Tr}) into Eq. (\ref{Met1b}) we obtain
\begin{eqnarray}
ds^{2} &=&-\frac{r^{2}+r_{+}^{2}}{l^{2}}\left( \Xi dt-ad\phi \right) ^{2}+%
\frac{r^{2}dr^{2}}{(r^{2}+r_{+}^{2})f(r)}  \nonumber \\
&&+l^{2}f(r)\left( \frac{a}{l^{2}}dt-\Xi d\phi \right) ^{2}+\frac{%
r^{2}+r_{+}^{2}}{l^{2}}dX^{2},  \label{Metr2}
\end{eqnarray}
where $f(r)$ is the same as $f(r)$ given in Eq. (\ref{Fg2}). The gauge
potential is now given by
\begin{equation}
A_{\mu }=\frac{2}{(n-2)}\frac{ql^{(n-3)}}{(r^{2}+r_{+}^{2})^{(n-2)/2}}\left(
a\delta _{\mu }^{0}-l^{2}\Xi \delta _{\mu }^{\phi }\right).  \label{Pot2}
\end{equation}
The transformation (\ref{Tr}) generates a new metric, because it
is not a permitted global coordinate transformation \cite{Sta}.
This transformation can be done locally but not globally.
Therefore, the metrics (\ref{Met1b}) and (\ref{Metr2}) can be
locally mapped into each other but not globally, and so they are
distinct. Note that this spacetime has no horizon and curvature
singularity. However, it has a conical singularity at $r=0$. One
should note that these solutions are different from those
discussed in \cite {Deh1}, which were electrically charged
rotating black brane solutions in Gauss-Bonnet gravity. The
electric solutions have black holes, while the magnetic do not. It
is worthwhile to mention that this solution reduces to
the solution of Einstein-Maxwell equation introduced in \cite{Deh2} as $%
\alpha $ goes to zero.

\subsection{The general rotating longitudinal solution with more rotation
parameters}

For the sake of completeness we give the general rotating
longitudinal solution with more rotation parameters. The rotation
group in $n+1$ dimensions is $SO(n)$ and therefore the number of
independent rotation parameters is $[(n+1)/2]$, where $[x]$ is the
integer part of $x$. We now generalize the above solution given in
Eq. (\ref{Metr2}) with $k\leq \lbrack (n+1)/2]$ rotation
parameters. This generalized solution can be written as
\begin{eqnarray}
ds^{2} &=&-\frac{r^{2}+r_{+}^{2}}{l^{2}}\left( \Xi dt-{{\sum_{i=1}^{k}}}%
a_{i}d\phi ^{i}\right) ^{2}+f(r)\left( \sqrt{\Xi ^{2}-1}dt-\frac{\Xi }{\sqrt{%
\Xi ^{2}-1}}{{\sum_{i=1}^{k}}}a_{i}d\phi ^{i}\right) ^{2}  \nonumber \\
&&+\frac{r^{2}dr^{2}}{(r^{2}+r_{+}^{2})f(r)}+\frac{r^{2}+r_{+}^{2}}{%
l^{2}(\Xi ^{2}-1)}{\sum_{i<j}^{k}}(a_{i}d\phi _{j}-a_{j}d\phi _{i})^{2}+%
\frac{r^{2}+r_{+}^{2}}{l^{2}}dX^{2},  \label{Metr5}
\end{eqnarray}
where $\Xi =\sqrt{1+\sum_{i}^{k}a_{i}^{2}/l^{2}}$, $dX^{2}$ is the Euclidean
metric on the $(n-k-1)$-dimensional submanifold and $f(r)$ is the same as $%
f(r)$ given in Eq. (\ref{Fg2}). The gauge potential is

\begin{equation}
A_{\mu }=\frac{2}{(n-2)}\frac{ql^{(n-2)}}{(r^{2}+r_{+}^{2})^{(n-2)/2}}\left(
\sqrt{\Xi ^{2}-1}\delta _{\mu }^{0}-\frac{\Xi }{\sqrt{\Xi ^{2}-1}}%
a_{i}\delta _{\mu }^{i}\right) ;\hspace{0.5cm}{\text{(no sum on
\emph{i})}}. \label{Pot5}
\end{equation}
Again this spacetime has no horizon and curvature singularity. However, it
has a conical singularity at $r=0$. One should note that these solutions
reduce to those discussed in \cite{Deh1}.

\section{The Angular Magnetic Field Solutions\label{AngMag}}

In Sec. \ref{Lon} we found a spacetime generated by a magnetic
source which produces a longitudinal magnetic field along $x^{i}$
coordinates. In this section we want to obtain a spacetime
generated by a magnetic source that produce angular magnetic
fields along the $\phi ^{i}$ coordinates. Following the steps of
Sec. \ref{Lon} but now with the roles of $\phi $ and $x$
interchanged, we can directly write the metric and vector
potential satisfying the field equations (\ref{Geq})-(\ref{Str})
as
\begin{eqnarray}
ds^{2} &=&-\frac{r^{2}+r_{+}^{2}}{l^{2}}dt^{2}+\frac{r^{2}dr^{2}}{%
(r^{2}+r_{+}^{2})f(r)}  \nonumber \\
&&+(r^{2}+r_{+}^{2}){{\sum_{i=1}^{n-2}}}(d\phi^{i})^{2}+f(r)dx^{2},
\label{Metr3}
\end{eqnarray}
where $f(r)$ is given in Eq. (\ref{Fg2}). The angular coordinates $\phi ^{i}$%
's range in $0\leq \phi ^{i}<2\pi $. The gauge potential is now
given by
\begin{equation}
A_{\mu }=-\frac{2}{(n-2)}\frac{q l^{(n-2)}}{(r^{2}+r_{+}^{2})^{(n-2)/2}}%
\delta _{\mu }^{x}.  \label{Pot3}
\end{equation}
The Kretschmann scalar does not diverge for any $r$ and therefore
there is no curvature singularity. The spacetime (\ref{Metr3}) is
also free of conic singularity. In addition, it is notable to
mention that the radial geodesic passes through $r=0$ (which is
free of singularity) from positive values to negative values of
the coordinate $r$. This shows that the radial coordinate in Eq.
(\ref{Metr3}) can take the values $-\infty <r<\infty $. This
analysis may suggest that one is in the presence of a traversable
wormhole with a
throat of dimension $r_{+}$. However, in the vicinity of $r=0 $, the metric (%
\ref{Metr3}) can be written as

\begin{eqnarray}
ds^{2} &=&-\frac{r_{+}^{2}}{l^{2}}dt^{2}+2\frac{l^{2}}{r_{+}^{2}}\left( n+%
\frac{4q^{2}l^{2n-2}}{(n-1)r_{+}^{2n-2}}\right) ^{-1}dr^{2}  \nonumber \\
&&+r_{+}^{2}d\Omega ^{2}+\frac{1}{2l^{2}}\left( n+\frac{4q^{2}l^{2n-4}}{%
(n-1)r_{+}^{2n-2}}\right) r^{2}dx^{2},
\end{eqnarray}
which clearly shows that, at $r=0$, the $x$ direction collapses
and therefore we have to abandon the wormhole interpretation.

To add linear momentum to the spacetime, we perform the boost
transformation $[t\mapsto \Xi t-(v/l)x$, $x\mapsto \Xi x-(v/l)t$]
in the $t-x$ plane and obtain
\begin{eqnarray}
ds^{2} &=&-\frac{r^{2}+r_{+}^{2}}{l^{2}}\left( \Xi dt-\frac{v}{l}dx\right)
^{2}+f(r)\left( \frac{v}{l}dt-\Xi dx\right) ^{2}  \nonumber \\
&&+\frac{r^{2}dr^{2}}{(r^{2}+r_{+}^{2})f(r)}+(r^{2}+r_{+}^{2})d\Omega ^{2},
\label{Metr4}
\end{eqnarray}
where $v\ $is a boost parameter and $\Xi =\sqrt{1+v^{2}/l^{2}}$. The gauge
potential is given by

\begin{equation}
A_{\mu }=\frac{2}{(n-2)}\frac{\lambda l^{(n-3)}}{(r^{2}+r_{+}^{2})^{(n-2)/2}}%
\left( v\delta _{\mu }^{0}-l\Xi \delta _{\mu }^{x}\right).  \label{Pot4}
\end{equation}
Contrary to transformation (\ref{Tr}), this boost transformation
is permitted globally since $x$ is not an angular coordinate. Thus
the boosted solution (\ref{Metr4}) is not a new solution. However,
it generates an electric field.

\section{ The Conserved Quantities of A Magnetic Rotating Brane\label{Conserv}}

It is well known that the gravitational action given in Eq.
(\ref{Actg}) diverges. A systematic method of dealing with this
divergence in Einstein gravity is through the use of the
counterterms method inspired by the anti-de Sitter conformal field
theory (AdS/CFT) correspondence \cite{Mal}. This conjecture, which
relates the low energy limit of string theory in asymptotically
anti de-Sitter spacetime and the quantum field theory living on
the boundary of it, have attracted a great deal of attention in
recent years. This equivalence between the two formulations means
that, at least in principle, one can obtain complete information
on one side of the duality by performing computation on the other
side. A dictionary translating between different quantities in the
bulk gravity theory and their counterparts on the boundary has
emerged, including the partition functions of both theories. This
conjecture is now a fundamental concept that furnishes a means for
calculating the action and conserved quantities intrinsically
without reliance on any reference spacetime \cite{Sken1,BK,Od1}.
It has also been applied to the case of black holes with constant
negative or zero curvature horizons \cite{Deh3} and rotating
higher genus black branes \cite {Deh4}. Although the AdS/CFT
correspondence applies for the case of a specially infinite
boundary, it was also employed for the computation of the
conserved and thermodynamic quantities in the case of a finite
boundary \cite {Deh5}. The counterterm method has also been
extended to the case of asymptotically de Sitter spacetimes
\cite{dS}.

All of the work mention in the last paragraph was limited to
Einstein gravity where the universal and widely accepted
Gibbons-Hawking boundary term \cite{Gib} is known. In Einstein
gravity the Gibbons-Hawking term, which is the trace of the
extrinsic curvature of the boundary, will be added to the action
in order to have a \thinspace well-defined variational principle.
The main difference of higher derivative gravity from Einstein
gravity is that the boundary term which make the variational
principle well-behaved is much more complicated \cite{Od2,Fuc}.
However, the boundary term that makes the variational principle
well behaved is known for the case of Gauss-Bonnet gravity
\cite{Mye, Dav}. Here we want to apply the counterterm method to
the case of magnetic brane solutions introduced in this paper. In
order to do this we choose the following counterterm
\begin{equation}
I_{b}=I_{b}^{(1)}+I_{b}^{(2)},  \label{metc}
\end{equation}
where $I_{b}^{(1)}$ is given as \cite{Mye,Dav}
\begin{equation}
I_{b}^{(1)}=\int_{\delta \mathcal{M}}d^{n}x\sqrt{-\gamma }\left\{ K+2\alpha %
\left[ \frac{1}{n}\left( +3KK_{cd}K^{cd}-2K_{ac}K^{cd}K_{d}^{\ a}-K^{3}\right)-2%
\widehat{G}^{ab}K_{ab} \right] \right\} .  \label{Ib1}
\end{equation}
In Eq. (\ref{Ib1}), $K$ is the trace of the extrinsic curvature
$K_{ab}$ of any boundary $\partial \mathcal{M}$ of the manifold
$\mathcal{M}$, with induced metric $\gamma _{ab}$, and
$\widehat{G}^{ab}$ is the $n$-dimensional Einstein tensor
corresponding to  the metric $\gamma _{ab}$. The second boundary
term $I_{b}^{(2)}$ is the counterterm that cancels the divergence
appearing in the limit of $r\rightarrow \infty $. This term is
given only in terms of the boundary quantities which do not affect
the variational principle. Note that for our
spacetimes the Riemann curvature of the boundary is zero and therefore $%
I_{b}^{(2)}$ has only one term given as
\begin{equation}
I_{b}^{(2)}=-\int_{\delta \mathcal{M}}d^{n}x\sqrt{-\gamma }\left( \frac{n-1}{%
l_{\mathrm{eff}}}\right) ,  \label{Ib2}
\end{equation}
where $l_{\mathrm{eff}}$\ is given in Eq. (\ref{leff}). One may
note that this counterterm has exactly the same form as the
counterterm in Einstein gravity for a spacetime with zero
curvature boundary in which $l$ is replaced by
$l_{\mathrm{eff}}$.The total action can be written as a linear
combination of the gravity term (\ref{Actg}), and the counterterms
(\ref{Ib1}) and (\ref{Ib2})
\begin{equation}
I=I_{g}+I_{b}.  \label{Itot}
\end{equation}
It is worthwhile to mention that the action (\ref{Itot}) has no
$r$ divergence in various dimensions for spacetimes with zero
curvature boundary. However, the boundary term introduced in
\cite{Od2}, does not have this property. In fact, the coefficients
of various terms in the action of \cite{Od2} in different
dimension should be chosen different, in order to remove the
$r$-divergence of the action. Having the total finite action, one
can use the Brown-York definition of stress-energy tensor
\cite{Brown} to construct a divergence-free stress-energy tensor.
Note that the last term in Eq. (\ref{Ib1}) is zero, for our case,
and therefore one may write
\begin{eqnarray}
T^{ab} &=&\sqrt{-\gamma }\{(K\gamma ^{ab}-K^{ab}) \\
&&+\frac{2\alpha }{n}\left(
3K^{2}K^{ab}-3K_{cd}K^{cd}K^{ab}+a_{n}KK^{ac}K_{c}^{\
b}+b_{n}K^{ac}K_{cd}K^{db}\right)
\nonumber \\
&&+\frac{2\alpha }{n}\left(
3KK_{cd}K^{cd}-2K_{ac}K^{cd}K_{db}-K^{3}\right) \gamma
^{ab}-\left(\frac{n-1}{l_{\mathrm{eff}}}\right)\gamma ^{ab}\}.
\label{Stres}
\end{eqnarray}
One may note that when $\alpha $ goes to zero, the stress-energy tensor (\ref
{Stress}) reduces to that of Einstein gravity.

The conserved quantity associated with a Killing vector $\xi ^{a}$
is
\begin{equation}
\mathcal{Q}(\xi )=\int_{\mathcal{B}}d^{n}x\sqrt{\sigma }u^{a}T_{ab}\xi ^{b},
\label{Con}
\end{equation}
where $\mathcal{B}$ is the hypersurface of fixed $r$ and $t$,
$u^a$ is the unit normal vector on $\mathcal{B}$, and $\sigma $ is
the determinant of the metric $\sigma _{ij}$, appearing in the
ADM-like decomposition of the boundary metric,
\begin{equation}
ds^{2}=-N^{2}dt^{2}+\sigma _{ij}(dx^{i}+N^{i}dt)(dx^{j}+N^{j}dt).
\label{ADM}
\end{equation}
In Eq. (\ref{ADM}), $N$ and $N^{i}$ are the lapse and shift
functions,
respectively. For the spacetimes introduced in this paper, the $(n-1)$%
-dimensional boundaries $\mathcal{B}$ have timelike Killing vector
($\xi =\partial /\partial t$), rotational Killing vector $(\zeta
_{i}=\partial /\partial \phi ^{i})$, and translational Killing
vector $(\varsigma _{i}=\partial /\partial x^{i})$. Thus, one
obtains the conserved mass,
angular momentum, and linear momentum of the system enclosed by the boundary $%
\mathcal{B}$ as
\begin{eqnarray}
&& M =\int_{\mathcal{B}}d^{n-1}x\sqrt{\sigma }T_{ab}u^{a}\xi ^{b},
\label{Mas} \\
&& J_{i} =\int_{\mathcal{B}}d^{n-1}x\sqrt{\sigma }T_{ab}u^{a}\zeta
_{i}^{b},
\label{Amom} \\
&& P_{i} =\int_{\mathcal{B}}d^{n-1}x\sqrt{\sigma
}T_{ab}u^{a}\varsigma _{i}^{b}.  \label{Lmom}
\end{eqnarray}

We now apply this counterterm method to the case of
five-dimensional spacetimes (\ref{Metr2}) and (\ref{Metr4}). It is
easy to verify that the $r$ divergence of the action is removed by
the counterterm (\ref{Ib2}). The
divergence terms of the mass of the spacetimes (\ref{Metr2}) and (\ref{Metr4}%
) in five dimensions will be removed if one choose $a_{4}=-4$ and $b_{4}=-8$%
. Using these coefficients together with Eqs. (\ref{Mas}) and (\ref{Amom}),
the mass and the angular momentum densities of the spacetime (\ref{Metr2})
in five dimensions can be calculated as
\begin{eqnarray}
&& M =m\left[ 4(\Xi ^{2}-1)+1\right] ,  \label{M2} \\
&& J =-16\Xi ma.
\end{eqnarray}
The mass of the spacetime (\ref{Metr4}) in five dimensions is the same as (%
\ref{M2}), its angular momentum is zero, and its linear momentum
is
\begin{equation}
P=-16\Xi mv.  \label{P4}
\end{equation}
Of course, one can apply this method to compute the mass, and
angular and linear momenta of the spacetime in various dimensions.

Next, we calculate the electric charge of the solutions. To
determine the electric field we should consider the projections of
the electromagnetic field tensor on special hypersurfaces. The
normal to such hypersurfaces for spacetimes with longitudinal
magnetic field is
\begin{equation}
u^{0}=\frac{1}{N},\text{ \ }u^{r}=0,\text{ \ }u^{i}=-\frac{N^{i}}{N},
\end{equation}
and the electric field is $E^{\mu }=g^{\mu \rho }F_{\rho \nu
}u^{\nu }$. Then the electric charge density $Q$ of the spacetimes
(\ref{Metr2}) and (\ref {Metr4}) can be found by calculating the
flux of the electromagnetic field at infinity, yielding
\begin{equation}
Q=\frac{1}{4\pi }\sqrt{\Xi ^{2}-1}q.  \label{elecch}
\end{equation}
Note that the electric charge density is proportional to the
rotation parameter or boost parameter and is zero for the case of
static solutions.

\section{CLOSING REMARKS}

In this paper, we added the Gauss-Bonnet term to the
Einstein-Maxwell action with a negative cosmological constant. We
introduced two classes of solutions which are asymptotically
anti-de Sitter. The first class of solutions yields an
$(n+1)$-dimensional spacetime with a longitudinal magnetic field
[the only nonzero component of the vector potential is $A_{\phi
}(r)$] generated by a static magnetic brane. We also found the
rotating spacetime with a longitudinal magnetic field by a
rotational boost transformation. We found that these solutions
have no curvature singularity and no horizons, but have conic
singularity at $r=0$. In these spacetimes, when all the rotation
parameters are zero (static case), the electric field vanishes,
and therefore the brane has no net electric charge. For the
spinning brane, when one or more rotation parameters are nonzero,
the brane has a net electric charge density which is proportional
to the magnitude of the rotation parameter given by $\sqrt{\Xi
^{2}-1}$. The second class of solutions yields a spacetime with
angular magnetic field. These solutions have no curvature
singularity, no horizon, and no conic singularity. Again, we found
that the branes in these spacetimes have no net electric charge
when all the boost parameters are zero. We also showed that, for
the case of traveling branes with nonzero boost parameter, the net
electric charge density of the brane is proportional to the
magnitude of the velocity of the brane ($v$).

The counterterm method inspired by the AdS/CFT correspondence
conjecture has been widely applied to the case of Einstein
gravity. Here we applied this method to the case of Gauss-Bonnet
gravity and calculated the conserved quantities of the two classes
of solutions. We found that the counterterm (\ref{Ib2}) has only
one term, since the boundaries of our spacetimes are
curvature-free. Other related problems such as the application of
the counterterm method to the case of solutions of higher
curvature gravity with nonzero curvature boundary remain to be
carried out.

\end{document}